# FABRICATION OF SURFACE-ENHANCED RAMAN SPECTROSCOPY SUBSTRATES USING SILVER NANOPARTICLES PRODUCED BY LASER ABLATION IN LIQUIDS

Annah M. Ondieki[1*], Zephania Birech[**,1], Kenneth A. Kaduki[1] & Peter W. Mwangi[2]

[1] Department of Physics, University of Nairobi, P.O Box 30197-00100, Nairobi, Kenya, [2] Department of Medical Physiology, University of Nairobi, P.O Box 30197-00100, Nairobi, Kenya

Corresponding Authors: [*] moraa94annah@gmail.com (OA); ** birech@uonbi.ac.ke (BZ);

**Abstract**

This research describes the use of surface-enhanced Raman spectroscopy (SERS) substrates based on colloidal silver nanoparticles (AgNPs) produced by laser ablation of silver granules in pure water that are inexpensive, easy to make, and chemically stable. Here, the effects of the laser power, pulse repetition frequency, and ablation duration on the Surface Plasmon Resonance peak of AgNPs solutions, were used to determine the optimal parameters. Also, the effects of the laser ablation time on both ablation efficiency and SERS enhancement were studied. The synthesized AgNPs were characterized by UV-Vis spectrophotometer and Raman spectrometer. The Surface Plasmon Resonance peak of AgNP solutions was centered at 404 nm confirming their synthesis. Using Raman spectroscopy, they had main bands centered at 196 cm$^{-1}$ (O=Ag$_2$/Ag-N stretching vibrations), 568 cm$^{-1}$ (N-H out of plane bending); 824 cm$^{-1}$ (symmetric deformation of the NO$_2$); 1060 cm$^{-1}$ (N-H out of plane bending); 1312 cm$^{-1}$ (symmetric stretching of NO$_2$); 1538 cm$^{-1}$ (N-H in-plane bending); and 2350 cm$^{-1}$ (N$_2$ vibrations). Their Raman spectral profiles remained constant within the first few days of storage at room temperature implying chemical stability. The Raman signals from blood were enhanced when mixed with AgNPs and this depended on colloidal AgNPs concentration. Using those generated by 12 hrs ablation time, an enhancement of 14.95 was achieved. Additionally, these substrates had an insignificant impact on the Raman profiles of samples of rat blood when mixed with them. The Raman peaks noted were attributed to C-C stretching of glucose (932 cm$^{-1}$); C-C stretching of Tryptophan (1064 cm$^{-1}$); C-C stretching of β Carotene (1190 cm$^{-1}$); CH$_2$ wagging of proteins (1338 and 1410 cm$^{-1}$); carbonyl stretch for proteins (1650 cm$^{-1}$); C≡N vibrations for glycoproteins (2122 cm$^{-1}$). These SERS substrates can be applied to areas such as forensics to distinguish between human and other animal blood, monitoring of the efficacy of drugs, disease diagnostics such as diabetes, and pathogen detection. All this can be achieved by comparing the Raman spectra of the biological samples mixed with the synthesized SERS substrates for different samples. Thus, the results on the use of inexpensive, simple-to-prepare Raman substrates have the possibility of making surface-enhanced Raman spectroscopy available to laboratories with scarce resources in developing nations.

**1. Introduction**

The Surface-Enhanced Raman spectroscopy (SERS) approach using the plasmonic characteristics of metallic nanostructures considerably improves the Raman signal [1], especially for biological samples. Commercially available SERS substrates such as calcium fluoride are expensive [2]. Thus, metallic nanoparticles such as silver, gold, and copper are often used to create them. Silver nanoparticles (AgNPs) are increasingly preferred over gold and copper because they are cost-effective with great sensitivity and unique optical, physical, catalytic, and chemical capabilities [1]. Physical and chemical procedures are used to create them [3]. Chemical procedures are dangerous and produce nanoparticles with solvent contamination [4], whereas physical methods produce nanoparticles with uniformly distributed sizes and of the best purity [5]. Laser ablation, laser pyrolysis, and electro-spraying are some of the physical approaches that have been used to synthesize AgNPs [3,6]. Among these methods, laser ablation in liquids has gained popularity due to its ease of use, inexpensive, capacity to produce chemically stable NPs without contamination, and a higher degree of control over the density and size of particles [5,7].

Surface-plasmon resonance (SPR) dominates the optical absorption spectra of metal nanoparticles, which shift to longer wavelengths with increasing particle size [8]. These plasmon resonances are caused by the collective electron's oscillations that are sensitive to size, shape, and the surrounding medium. The UV-Visible absorbance is used to detect the presence of AgNPs because they have plasmon resonances in the visible range. This is reported to occur around 400 nm, with the wavelength varying depending on the nanoparticle size and shape [9]. Raman spectroscopy is also used to characterize AgNPs by examining their molecular fingerprint [10]. Despite the significance of controlling particle size distribution to produce high-quality SERS substrates, little is known about how laser ablation in liquids produces colloid particles or how this affects the amplification of Raman spectra in biological samples. As a result, this research examines the impact of AgNP concentration on SERS spectra as well as the ablation time on ablation efficiency. Here, the background signals emitted by SERS substrates generated via laser ablation were insignificant to the Raman signals of blood samples mixed with them. When stored at room temperature, they were chemically stable within the first few days and improved biological sample Raman signals when used. This enhancement improved with the concentration of AgNPs. Thus, unlike, commercially accessible SERS substrates, those based on the ablation of silver granules in pure water are cost-effective, pure, easy to make, and chemically stable.

## 2. Experimental Setup

It was crucial to establish the ideal ablation parameters because the laser energy, ablation time, and pulse repetition frequency (PRF) influence the concentration of the NPs generated [11,12]. To do this, the AgNP colloids were prepared by laser ablation in liquid (distilled water) by changing one parameter while keeping other parameters constant. Here, first, the laser Pulse Energy was varied in the range of 100 – 250 mJ keeping the PRF and ablation duration constant. Then, ablation duration varied in the range of 10-180 minutes keeping pulse energy and PRF constant. Finally, the PRF varied from 1 to 15 Hz keeping the laser pulse energy and ablation duration constant. The 99.99% pure silver granules (Sigma-Aldrich) placed on the bottom of a glass beaker containing 5 ml of distilled water (see figure 1) were ablated by a Q-Switched Nd:YAG laser emitting 8 ns long pulses at 1064 nm wavelength resulting in AgNPs suspension in water.

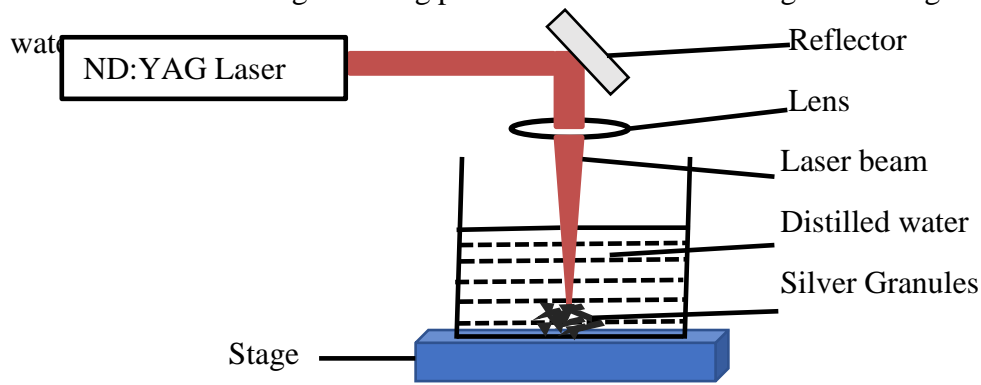

*Figure 1: Laser Ablation of silver nanoparticles*

The synthesized nanoparticles (AgNPs) placed in quartz cuvettes were then characterized optically by measuring their absorption spectrum using UV-Vis Spectrophotometer (solidspec-3700

DUV230, A11094500005). The appropriate parameters were those used to generate AgNP colloids with the highest absorbance (see figure 2). The AgNPs generated by 5 PRF, 250 mJ pulse energy, and 0.5-12 hours ablation time were used in this study. Their concentrations were deduced (at the center peak maximum of the surface plasmon resonance, SPR, band) using the Beer-Lambert law, $c = \frac{A}{\epsilon d}$, where c is the concentration of NPs, A is the absorbance, d is the cuvette path length and $\epsilon$ is the extinction coefficient ($145 \times 10^{10}\ L\ mol^{-1} cm^{-1}$ at 405.6 nm [13]). On using the values given in [13] for particle size at different absorption maxima wavelengths, we deduced that the particle sizes in this work were about 30 nm in diameter at 404 nm absorption maximum (see Fig. 3.a). Raman measurements of the AgNPs (on Al foil-wrapped glass slide) generated by 0.5-12 hours ablation time were done to confirm the increase concentration with ablation time (see figure 3.b). The chemical stability of the NPs was examined by measuring their (AgNPs drop on Al foil-wrapped glass slide) Raman spectra (taken at ten random spots per a drop) on different days (day 1 to day 30) after synthesis and storage at room temperature. The portable Raman spectrometer (EZRaman- N Portable analyzer system, Enwave Optronics, Inc.) equipped with a 785 nm laser (~300 mW), fiber-optic probe (7 mm working distance), charge-coupled detector (cooled at -25$^0$C), and spectrograph (f/1.6) (EZRaman-NP-785) was used in this work.

For studying the biological fluid samples, whole blood drawn from Sprague Dawley (SD) rats were used. These animals were kept at the department of medical physiology, University of Nairobi in conditions as described elsewhere [14]. Blood extractions were done at the same department (medical physiology) and supplied to the physics department where they were refrigerated at 4ºC. All necessary ethical approvals were sought and granted. For the SERS experiment, the AgNPs (about 90 µL from each concentration) were first mixed with the blood samples (about 10 µL) then smeared, separately, on a clean aluminum foil-wrapped microscope glass slide, and on a clean glass slide, let to air dry for one hour, and later each sample excited with a 785 nm laser. Also, a thin smear (about 10 µL) and a thick smear of blood (about 50 µL) were dried on both aluminum-wrapped glass slides and on a clean glass slide and excited. Raman spectral data from 12 random spots on each sample were recorded and analyzed. The thin smear guaranteed significant SERS activity. To compare the amount of Raman signal amplification produced by using the colloidal AgNPs, a thin blood smear was made on both the clean Al foil and glass slide, and a similar thin spread was done after mixing blood with colloidal AgNPs. Raman spectra from the clean Al foil and glass were also obtained. Raman spectra from each were collected, averaged, plotted on the same axes, and the area under the curve (AUC) was calculated. After that, the enhancement factor (E.F) was calculated as [2]. The Raman experimental parameters used were as follows: excitation power at the sample position, ~150 mW; exposure time, 5 s; spectral averaging, 5 s; boxcar, 1, and fiber-coupled laser output (~100 μm, 0.22 NA). Spectral data pre-processing involved mean centering, and background correction. This was done in MATLAB R2021a (version 9.10.0.1602886, The MathWorks Inc., Natick, MA) scripting environment. For plotting, ORIGIN software (OriginPro 2021, version 9.8.0.200) was used.

## 3. Results and Discussion

### 3.1. Effect of laser Energy, laser Ablation time and PRF on SPR spectrum

Synthesis of significant amounts of AgNP colloids implicates high SERS enhancements. High concentrations of these colloids generated via laser ablation are dependent on the laser ablation parameters such as laser energy, PRF, and ablation time. Figure 2 shows the absorption spectra of the produced NPs solutions as well as their absorbance average band intensities.

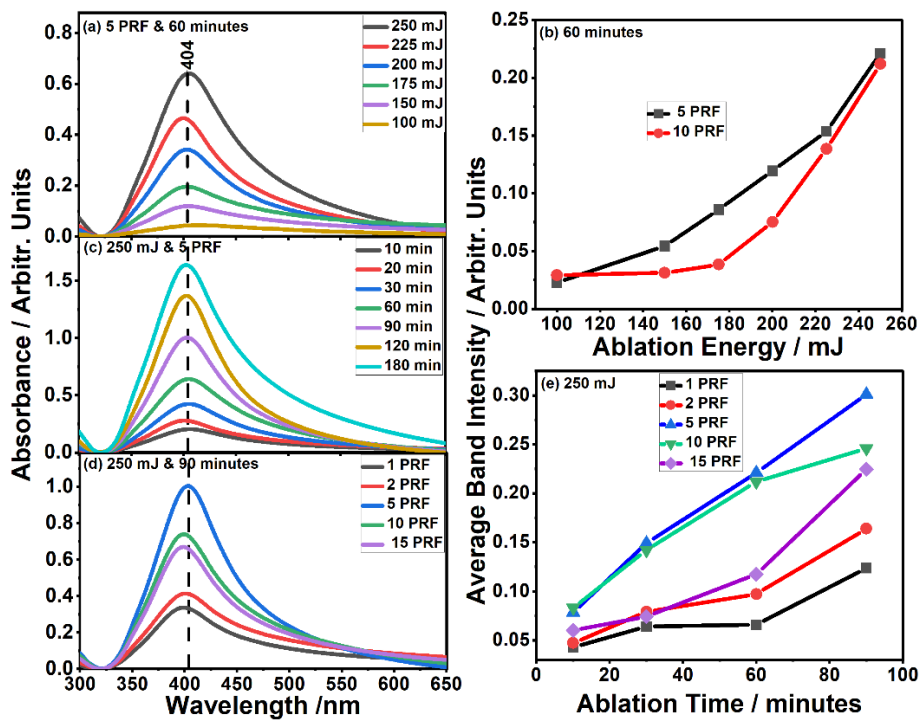

Figure 2: *UV-Vis absorption spectra for AgNPs solutions synthesized via laser ablation using (a) different laser ablation energies with constant PRF and Time and their (b) average SPR band*

*intensity, and (c) varying ablation time at fixed energy and PRF; (d) varying PRF at fixed ablation time and Energy and their (e) average SPR band intensity.*

The absorption peak was noted at around 404 nm in all samples due to Surface plasmon resonances (SPR) (see figure 2(a), 2(c), 2(d)) implying AgNPs were indeed generated [9]. The consistency in the SPR peak position and bandwidth (404 nm) confirmed the comparability of the size distribution of the NP solutions [15]. The intensity of the absorption peaks varied depending on the laser ablation energy. The intensity of absorption peaks increased with increasing the laser pulse energy which suggested that there is an increase in NPs concentration as the laser pulse energy increases [16,17]. Since the highest energy i.e 250 mJ, produced the largest absorbance, it was then used to obtain the appropriate PRF and ablation time to use. Using the highest energy (250 mJ), it is seen that there is an upward shift of the SPR peak in the AgNP solutions indicating a rapid increase in the number of NPs fabricated as ablation time is increased (see Figures 2(c)). Figure 2(d), shows the absorption spectra for AgNPs solutions synthesized by 250 mJ laser ablation energy and 90 minutes ablation time varying PRF. The smallest absorbance corresponded to the AgNPs solutions synthesized by the PRF of 1 Hz followed by 2 Hz, 15 Hz, 10 Hz, and 5 Hz respectively. Though 15 Hz is the highest PRF, it did not give the largest absorbance. This reduction in the SPR peak intensity can be attributed to more heat in the target (Ag granules) due to reheating by the successive laser pulses. The excess heat input caused the molten zone under the ablated area to expand instead of amplifying the vaporization [16]. The decrease in the vaporization process caused a decrease in the NPS formation rate thus a downward shift in the SPR peak. Therefore, the optimum PRF to use in ablating Ag granules was noted to be 5 Hz. In addition, Figure 2(e), showed that absorbance increased with an increase in ablation time for all PRFs implying the increase in the concentration of NPs as ablation time increased.

### 3.2. AgNPs and SERS substrate characterization

The optical properties of AgNPs are known to change depending on particle shape [18], size [13], composition, and surface capping [4]. The particles synthesized in this work were found to display a maximum absorption band centered at around 404 nm (as displayed in Fig. 3a) due to Surface plasmon resonances (SPR) [9]. As expected, the band intensity increased as the ablation time was increased signifying elevated concentrations of the nanoparticles (see Fig. 3c) [19]. Using the extinction coefficient in [13] (for NP size 30 nm) to calculate the AgNPs concentration in distilled water, figure 3c was plotted and curve fitting done using a parabola fit. The color of the formed colloids also varied from light yellow to dark brown (see inset, Fig. 3a).

To investigate further, Raman analysis was done on the samples. This is because it is known that the intensity of Raman spectral bands correlates to the concentration of molecules under study in a sample [20]. Here, Raman spectroscopy was used to first confirm the concentration dependence on ablation time and investigate the chemical stability of the synthesized NPs. Indeed, the intensity of the Raman bands in the spectra obtained from AgNPs as displayed in Fig. 3b was observed to increase with ablation time. As expected, the area under curve (AUC) values of the Raman spectra from the NPs increased with ablation time (see Fig 3d). The Raman spectra of distilled water used to synthesize AgNPs and AgNPs drop synthesized using 250 mJ, 5 Hz PRF with different ablation times are displayed in figure 3(b).

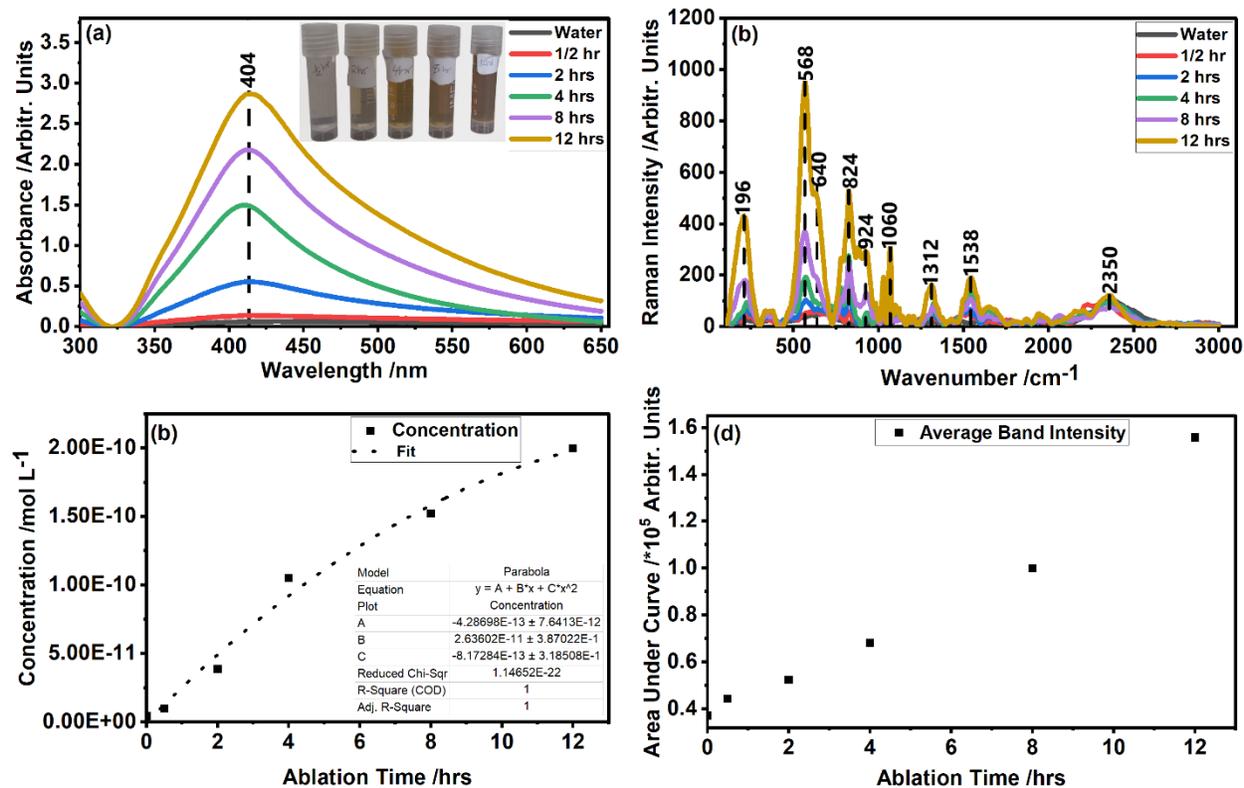

*Figure 3: Figure showing (a) absorbance spectra; (b) Raman spectra; and (c) Concentration against ablation time spectra; and (d) Raman spectra area under the curve (AUC) against ablation time spectra for AgNPs synthesized via 5 PRF, 250 mJ and 0.5 to 12 hours ablation time.*

Given that Raman spectra depend on how light interacts with chemical bonds in a substance, it is used to characterize colloidal AgNPs [10]. The Raman spectral profiles of AgNPs synthesized using 2, 4, 8 and 12 hours are fairly identical with different additional bands when compared with the spectra of water and those synthesized using lower ablation time (0.5 hour). The Raman bands seen are those centered at 196 cm$^{-1}$ (O=Ag$_2$/Ag-N stretching vibrations) [21,22]; 568 cm$^{-1}$ (N-H out of plane bending) [23]; 824 cm$^{-1}$ (symmetric deformation of the NO$_2$) [24]; 1060 cm$^{-1}$ (N-H out of plane bending) [23]; 1312 cm$^{-1}$ (symmetric stretching of NO$_2$)[24]; 1538 cm$^{-1}$ (N-H in-plane bending) [25]; and 2350 cm$^{-1}$ (N$_2$ vibrations) [26]. It was established that the vibrational signature attributable to Ag was that one centered around 196 cm$^{-1}$ and other bands were seen due to the presence of nitrogen in the air since ablation was done on air, oxygen and hydrogen from distilled water.

Silver is known to readily oxidize in air, thus Raman substrates based on it are frequently unstable [2]. This makes it necessary to look at the chemical stability of the AgNPs created via laser ablation in water. The Raman spectra of AgNPs synthesized using 12 hours ablation time were taken for 3 different days (1$^{st}$, 14$^{th,}$ and 30$^{th}$) after synthesis (see fig. 4).

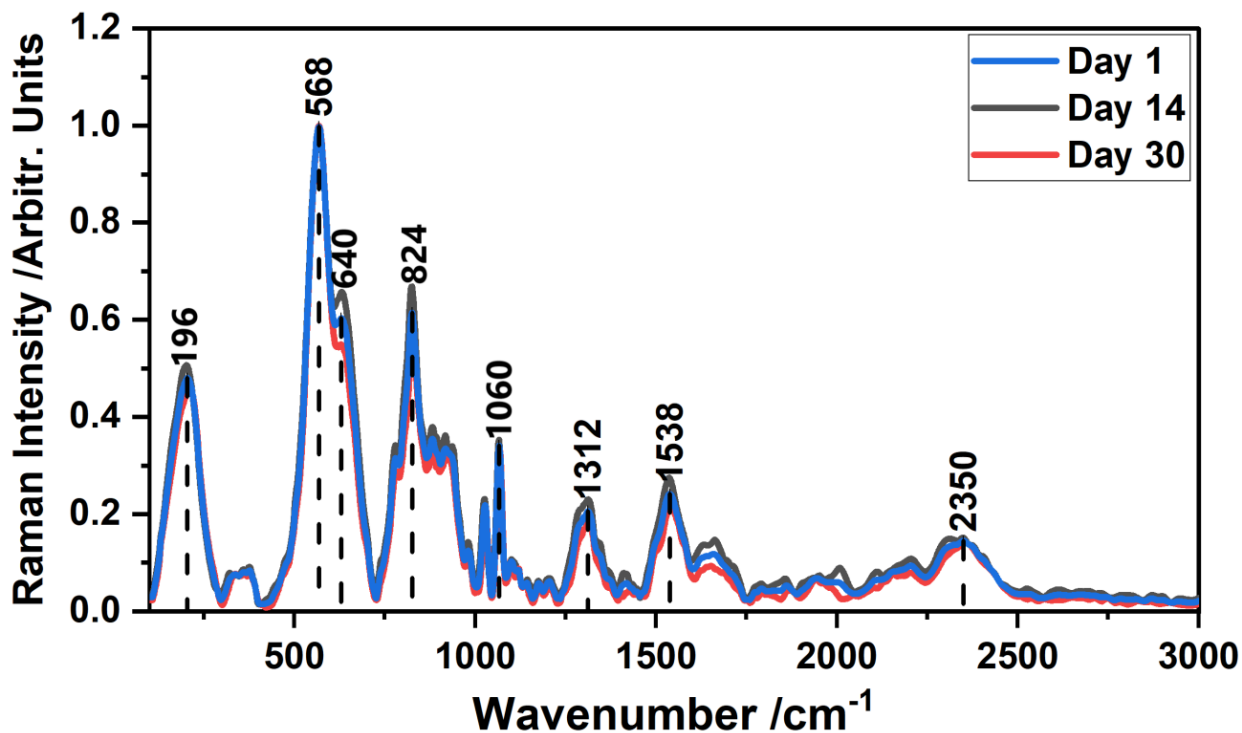

*Figure 4: Figure showing that the synthesized AgNPs were chemically stable as the Raman profiles were invariant several days after synthesis.*

The spectral profiles for the three days (1, 14 and 30) were noted to be identical suggesting chemical stability. The prominent bands were noted to be centered around 196, 568, 824, 1060,

1312, and 1538 cm$^{-1}$. These bands were consistent throughout all days, indicating that the molecules and bonds that were related to these signals never changed over the first month [2]. SERS substrates that are chemically stable and appropriate for room temperature application have been successfully created using AgNPs synthesized by laser ablation.

**3.3: AgNPs as SERS substrates for whole blood studies**

Raman signals from biological samples such as whole blood are often suppressed by autofluorescence signals from both the samples and from the glass substrates [2]. Figure 5 displays Raman spectra obtained from whole blood and a mixture of blood with different concentrations of AgNPs on the Aluminum foil (fig. 5a) as well as on a clean glass slide (fig. 5b). Also, shown are the Raman spectra of different concentrations of AgNPs generated via laser ablation (fig. 5c) and both blank Al wrapped glass slide and clean glass slide (fig. 5d).

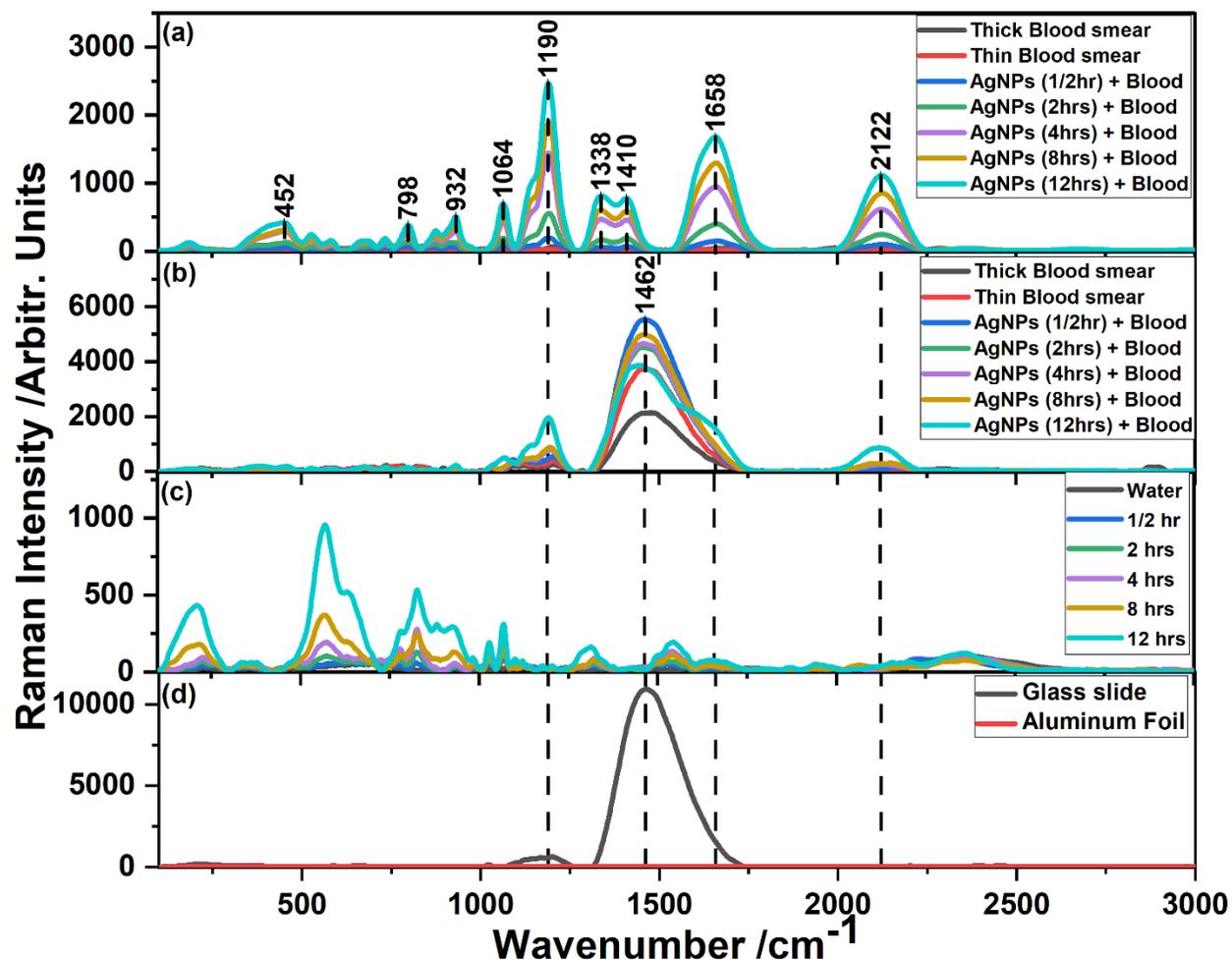

*Figure 5: Average Raman spectra obtained from the dry smear of SD rat's blood (thick and thin) and its mixture with different concentrations of AgNPs (generated via ablation) on a clean (a)*

*Aluminum wrapped glass slide and (b) glass slide. Also, Raman spectra of (c) AgNPs colloids synthesized generated via laser ablation in liquids using 5 Hz PRF, 250mJ and 0.5 to 12 hrs on Al Foil wrapped glass slide; and (d) both blank Al Foil and glass slide.*

Even though additional Raman bands (1190 and 2122 $cm^{-1}$) are seen on the spectra of the mixture (blood and AgNPs) dried on the glass slide when compared to those from blood on glass, the strong signals coming from the glass slide centered at roughly 1462 $cm^{-1}$ had a significant impact on both (see figure 5.b). On the other hand, the Raman signals from the blood dried on Al foil are very weak, unlike the signals from the mixture of blood and AgNPs. This is evident in figure 5(a) where prominent bands centered around 452, 798, 932, 1064, 1190, 1338, 1410, 1658, and 2122 $cm^{-1}$ are seen in the SERS spectra of blood with AgNPs dried on Al foil. Thus, the use of glass as a Raman substrate for biological sample analysis is discouraged because the spectral characteristics associated with blood were hardly discernible [2]. Costly substrates are frequently utilized, including sodium chlorite, calcium fluoride, and substrates with an aluminum coating. The aluminum foil did not influence the Raman signals from the blood sample [27], as seen in figure 5(a). As a result, it was discovered that while the SERS spectrum of blood could be seen on aluminum foil, it could hardly be seen on the glass slide. The Raman intensity of these bands increased with the use AgNPs generated using the longer ablation time. This implied that the longer the ablation time, the higher the concentration of AgNP colloids and hence the higher the enhancement of the Raman signal with the largest EF of 14.95. The Raman peaks were attributed to C-C stretching of glucose (932 $cm^{-1}$); C-C stretching of Tryptophan (1064 $cm^{-1}$) [28]; C-C stretching of β Carotene (1190 $cm^{-1}$) [28,29]; $CH_2$ wagging of proteins (1338 and 1410 $cm^{-1}$); carbonyl stretch for proteins (1650 $cm^{-1}$) [30]; C≡N vibrations for glycoproteins (2122 $cm^{-1}$) (Lin *et al.*, 2019; Mojica *et al.*, 2018) [31,32].

**Discussion**

For the initial characterization of produced NPs, UV-vis spectroscopy is a very efficient and reliable approach. Because of AgNPs' distinctive optical characteristics, they strongly interact with certain light wavelengths [33]. Electron motion is constrained when the diameter of these nanoparticles is smaller than or equal to the electron mean free path. Free electron band oscillates when an electromagnetic wave is incident on them. When the collective oscillation of the electrons in AgNPs is in resonance with the light wave, these free electrons give rise to an SPR absorption band (i.e. NPs strongly absorb) [34]. The absorption peak noted at around 404 nm in all samples implied that AgNPs were indeed generated [9]. The absorption peak intensity increased with higher ablation time suggesting that there is an increase in the number of NPs absorbing as the ablation time was increased [19]. This was also evident in the darkening of the AgNP colloids. The consistency in the SPR peak position (404 nm) confirmed the comparability of the size distribution of the NP solutions [13,15].

In Raman spectroscopy, the target molecules' vibration frequency determines the position of the peak. This enables the identification and specification of the substances in mixtures since each chemical bond has a unique vibration [10]. Ag-O vibration band centered at 196 cm$^{-1}$ in the Raman spectra (see figure 3b) served as confirmation for the formation of colloidal AgNPs via laser ablation in liquids. Since the Raman peak intensity correlates with the concentration of the substance identified, the upward shift of the Ag-O band as ablation time was increased implied an increase in the amount of AgNPs synthesized. This was also confirmed by Fig. 3d where the AUC for the Raman spectra of the colloidal AgNPs increased with ablation time. The other Raman bands noted were due to water molecules (for hydrogen and oxygen) and nitrogen since ablation was done in the air. Therefore, both absorbance and Raman spectroscopy confirmed the generation of AgNPs via laser ablation and the effect of laser ablation time on ablation efficiency. Once it was established that the longest ablation period resulted in the highest concentration, the chemical stability of the AgNPs produced by the 12-hour ablation time was examined. The Raman spectral profiles of the sample remained the same when days 1, 14 and 30 were plotted on one figure (see fig. 4) suggesting that the molecular bonds were not altered. This implied that these NPs are pure and don't need special storage within the first month, and are chemically stable, making them appropriate for usage in laboratories with limited resources.

The effect of ablation time was finally investigated on SERS activity. This is because it is well known that AgNPs improve the Raman signals of biological samples. As seen, biological materials (blood) mixed with synthesized SERS substrates allow for the observation of intense Raman signals with an E.F. of 14.95. These extremely strong Raman signals are due to the SERS effect which happens not only when the NP concentration is high, but also when the size of the metallic nanoparticles is equal to or less than the excitation light wavelength. These SERS substrates can be applied to areas such as forensics (distinguishing between human and other animal blood) [35], monitoring of the efficacy of drugs, disease diagnostics such as diabetes [36], and pathogen detection [37]. All this can be achieved by comparing the Raman spectra of the biological samples mixed with the synthesized SERS substrates for different samples. For instance, not only can cancerous samples be compared with non-cancerous and diabetic with un-diabetic but also a differentiation between various food-borne pathogens is possible as done elsewhere [36,37]

## 4. Conclusion

This research has demonstrated the potential of using pure colloidal AgNPs produced by laser ablation of silver granules in pure water as SERS substrates. It was discovered that the background signals emitted by these substrates had no impact on the Raman signals of blood samples that were mixed with them. They retained their chemical stability within the first few days when kept at room temperature making them suitable for use in laboratories with limited resources. As can be seen, adding blood to the created colloidal AgNP solutions enables the observation of strong Raman signals with an E.F of 14.95. These substrates' Raman spectral characteristics matched those of other groups' nanostructured SERS substrates exactly. To work with biological samples, colloidal AgNPs created by laser ablation can be utilized as inexpensive and pure Raman substrates.

**Acknowledgments**

We sincerely express our gratitude to Swedish International Development Cooperation Agency (SIDA), through the International Science Programme (ISP), Uppsala University, for sponsoring this research.